\documentstyle[pra,aps,preprint]{revtex}
\begin{document}
\draft
\title{
Field Induced Transition from 
Metal to Insulator \\ in the CMR Manganites
}
\author{Y. Endoh, H. Nojiri$^\dagger$, K. Kaneko$^\dagger$,  
K. Hirota, T. Fukuda$^{\ddagger}$, H. Kimura, \\ 
Y. Murakami$^\ast$ 
S. Ishihara$^\dagger$, S. Maekawa$^\dagger$,  
S. Okamoto$^\dagger$ and M. Motokawa$^\dagger$}
\address{CREST, Department of Physics, Tohoku University,  Sendai, 980-8578
Japan}
\address{$^\dagger$ CREST, Institute for Materials Research, Tohoku
University, 
Sendai, 980-8577 Japan}
\address{$^\ddagger$ SPring-8, JAERI, 1503 - 1 Kanaji, Kamigori-cho, Ako-gun, Hyogo 678-1200,
Japan}
\address{$^\ast$ CREST, Photon Factory, Institute of Materials Structure
Science,  KEK, Tsukuba 305-0801, Japan}
\date{October 11, 1998}
\maketitle
\begin{abstract}
The gigantic reduction of the electric resistivity under the applied magnetic 
field, CMR effect, is now widely accepted to appear in the vicinity of the 
insulator to metal transition of the perovskite manganites. Recently, we have 
discovered the first order transition from ferromagnetic metal to insulator 
in $\rm La_{0.88}Sr_{0.12}MnO_3$ of the CMR manganite. This phase transition induces the 
tremendous increase of the resistivity under the external magnetic field just 
near above the phase transition temperature. We report here fairly detailed 
results from the systematic experiments including neutron and synchrotron X-ray 
scattering studies.
\\
keywords: Metal Insulator Transition, CMR Manganites, Orbital Order,
Charge Order, Superexchange, Anisotropic Tensor Susceptibility,
Neutron scattering, Synchrotron x-ray scattering
\end{abstract}
\vfill
\eject
\narrowtext
\noindent
\section{Introduction}
The Metal - Insulator (MI) transition accompanied with the colossal magneto- 
resistance (CMR) effect in the manganese oxides (manganites) has been elucidated 
along the concept established in the studies for the strongly correlated electron 
systems \cite{r1}. Namely, $\rm LaMnO_3$ of the parent material of CMR is recognized as the 
Mott Insulator (Charge Transfer type like $\rm La_2CuO_4$). The "double exchange" 
interaction in these doped manganites was first proposed in 1950s \cite{r2}, but 
the presently discussed CMR effect cannot be comprehensive solely within this 
simple context of the "double exchange" mechanism \cite{r3}. Furthermore, unusual 
properties of the CMR effect are regarded as cooperative phenomena associated 
with the structural change due to a tiny atomic displacement, competing magnetic 
interactions and charge fluctuations between different valencies of manganese cations.
\par
In this respect, $\rm Mn^{3+}$ orbitals in $\rm LaMnO_3$ 
are Jahn-Teller ({\it JT}) active \cite{r4}, hence 
the degenerated $e_g$ "orbital" of $d$-electrons in the cubic crystalline field can 
be easily lifted  by lowering the crystal symmetry costing the lattice distortion 
energies. In fact, $\rm LaMnO_3$ is the antiferromagnetic insulator with the orthorhombic 
lattice structure (space group $Pnma$). Then the doping of holes by substituting 
trivalent La to divalent Sr cations promotes the simultaneous change in both 
magnetism and conductivity, which has been recognized to be originated by the 
"double exchange" mechanism \cite{r2}.  Since the CMR effect cannot be compatible to 
this simple mechanism, it has been believed that the CMR should result from the 
effect of the local distortion which is often defined as "polarons" \cite{r5}. The 
itinerancy of such polarons give rise to the conduction by either applying the 
magnetic field or decreasing temperature. However, there has been no definitive 
experimental result showing the existence of the polarons except speculative 
interpretation of data of electrical conductivity, mainly. Nonetheless, it is well 
recognized that the {\it JT} interaction plays a crucial role to determine the superexchange 
interactions in these transition metal oxides. 
\par
In this presentation, we focus on very unusual 
bulk properties in $\rm La_{1-x}Sr_xMnO_3$ 
($x \approx 0.12$) associated with not only the charge order (CO) but the "orbital" order 
(OO) and discuss that the CMR effect can be comprehensive by the role of the 
"orbital" degree of the freedom \cite{r6}. In order to investigate this subject in 
detail, we have conducted a decisive observation of the OO in the typical CMR 
manganites. Since this method is not familiar but very novel, we therefore 
briefly describe the principle of the experimental method \cite{r7}. Then we will 
argue that the electron correlations also act a significant key role for the OO and 
eventually  the CMR effect is relevant to the phase transition in the manganites. 
\par
The format of the paper is as follows. Bulk properties, in particular associated 
with the phase transition from ferromagnetic metal to insulator in the 
$ \rm La_{0.88}Sr_{0.12}MnO_3$ by either  lowering temperature 
and  applying the external field 
are presented. The experimental results of the OO will be followed by the 
description of the principle of detection of OO. Then the final section is 
devoted to discussions of our recent experimental results based on the recent 
novel concept of the "phase separation".
\section{Bulk Properties of $\rm La_{0.88}Sr_{0.12}MnO_3$}
$\rm La_{1-x}Sr_xMnO_3$ is recognized to be simplest among many CMR manganites and also 
it is the most extensively investigated material so far. Even so, the phase 
diagram of this system is already complicated near $x \approx 0.1$, where several phase 
boundaries of insulator-metal, lattice symmetry as well as magnetic structure 
are entangled with each other on the temperature ($T$) - doping concentration ($x$) 
diagram as shown in Fig. 1 \cite{r8}. In other words, if we understand correctly this 
complexity in the phase diagram, we might find a possible clue to the CMR mechanism 
in the manganites.
\par
The CO, lattice distortion and magnetic properties presumably couple with each 
other around $x \approx 0.125$ (1/8), but we must make it sure what extent and how. 
In order to elucidate this particular subject, it is important to grow a high 
quality single crystal. Now we all know that high quality single crystals can 
be grown by the conventional floating zone method using a lamp image focusing 
furnace. After characterizing this single crystal, which is determined to be 
$x=0.12$, we have systematically investigated both bulk and microscopic characters \cite{r9}. 
Here, we summarize the essence of our experiments showing an unusual phase transition 
from ferromagnetic metal to insulator transition by either applying magnetic field 
or reducing temperature \cite{r10}.
\par
First, we confirmed that sequential structural phase transitions occur varying 
temperature as shown in Fig. 2. Upon cooling from high temperatures, the crystal 
symmetry undergoes from pseudo-cubic to orthorhombic at $T_H = 291 K$. Then ferromagnetic 
long range order (lro) occurs below $T_C=170 K$ in the orthorhombic phase. The crystal 
symmetry changes again from orthorhombic to another pseudo cubic at $T_L=145 K$ 
accompanied with the jump in magnetic structure, which seems to be reentrant to 
the pseudo-cubic structure \cite{r8}, but in fact it is not so simple as described below.  
\par
According to high-resolution synchrotron  x-ray powder diffraction at NSLS in BNL, 
the determined crystal structure is more complicated [11]. The powder sample for 
this particular experiment was prepared by carefully crushing a small piece of the 
single crystal under acetone and was highly crystalline, with peak width 
of $\approx 0.01^\circ$, 
close to the instrumental resolution. The crystal structure above $T_H$ is not pseudo 
cubic but in fact orthorhombic ($Pnma$). Moreover, a splitting of some of the 
diffraction peaks was seen below $350 K$, which was confirmed to be a  single phase at 
400 K. Then both phases underwent reentrant-type transitions as just described above, 
the majority phase at $295 K$ and $145 K$, and the minority phase ($20 \%$) at 320 K and 
120 K, with $b/\sqrt{2} < a \approx c$ in the intermediate temperature 
region, characteristic of 
the {\it JT} ordered state. $20 \%$ of the minority phase was assigned to be the phase of 
relatively dilute Sr or hole concentration. The phase transition at $T_H$ is characterized 
by the abrupt decrease in the $b$ lattice parameter indicating the onset of the 
{\it JT} orbital ordering of $\rm LaMnO_3$ type. Then the crystal structure below $T_L$ was not 
well determined due to the considerable lowering of the symmetry, presumably triclinic 
($P2$). Note that the symmetry of the intermediate mixed phase is not orthorhombic 
but monoclinic ($\gamma \approx 90.1^\circ$). Though a further refinement 
is necessary, we have 
obtained rather important evidence of both phase separation and the change of the 
different crystal symmetry at above $T_H$ and below $T_L$ from this precise structural 
determination.
\par
Next, we summarize here the results of bulk measurements, in particular the magnetic 
properties under the external magnetic field \cite{r10}. Most of such experiments were 
performed at the high magnetic field laboratory of IMR, Tohoku university. 
We confirmed following important evidence of the excellent $\rm La_{0.88}Sr_{0.12}MnO_3$ 
single 
crystal from systematic magnetization measurements. The magnetization process shows 
a fairly weak anisotropy. This fact indicates that even below $T_L$, the canting of 
magnetic moment observed by neutron diffraction is not due to the major part of the 
single ion anisotropy, but presumably due to a weak anisotropic interaction by the 
lowering the crystal symmetry. Then there appears a hysteresis loop showing a jump 
in the magnetization process at each temperature between $T_L$ and $T_C$ in the 
intermediate phase. The transition field depends upon temperature, which increases 
monotonically with temperature. The hysteresis loop was confirmed to appear in both 
the megnetoresistance and magnetostriction measurements simultaneously. From the 
magnetoresistance measurements as shown in Fig. 3, we discovered very astonishing 
experimental evidence that the resistance abruptly increases at the transition 
field, which indicates the transition from metal to insulator at external fields 
in a certain temperature range of $T_L < T < T_C$. As already described above, the 
transition temperature increases as the increase of the magnetic field. Since the 
transition is associated with a clear jump in the magnetostriction, the 
metal-insulator phase transition is accompanied with the lattice distortion. Then the 
kinematical effect of the transition, or the relaxation effect in the magnetization 
in the field can well be interpreted by this fact accompanying the lattice distortion 
or the large magntostriction effect.
\par
According to the detailed analysis of the magnetization data, the jump at the 
transition field in magnetization is not due to the increase of the spontaneous 
magnetic moment, but due to the difference of the Curie temperature. In other words, 
the asymptotic Curie temperature of the lower temperature phase is higher than the 
actual $T_C$ (225 K). Then there appears another clear jump in magnetic susceptibility, 
$\chi$ at $T_H$, indicating the asymptotic Weiss temperature, where the 
$\chi^{-1} - T$ curve 
crosses zero, is about $205 K$ for the higher temperature phase. It means that the 
magnetic interaction does change in each phase with different crystal symmetry 
suggesting that  the interatomic interaction either superexchange or "double" 
exchange interaction is very sensitive to the atomic configuration.
\par
The latest measurements of the $x$ dependence of the magnetic properties 
around $x \approx 0.12$ show that the transitions in this range 
are very sensitive to $x$. 
Even changing $1 \%$ level in $x$ $(11 - 12.5 \%)$, the transition behavior changes 
significantly, which may reflect the change in the lattice structure in detail, 
though qualitatively no significant difference was reported so far.
\par
Now let's describe the results of neutron diffraction measurements. Above $T_L=145 K$, 
only (2,0,0) ferromagnetic component appears, which indicates a simple 3 dimensional 
(3D) isotropic ferromagnetic order below $T_C=170 K$. Below $T_L=145 K$, the very weak 
but distinct signal appears at (0,0,1) indicating that a tiny antiferromagnetically 
ordered component appears perpendicular to the (1,0,0) plane. Since the magnetization 
curves show no difference in crystal orientation, the net magnetic moment simply 
inclines from the $C$ axis by few degrees, probably $3 - 4^\circ$, 
so that the magnetic 
state is precisely the canted antiferromagnet, but the nature is ferromagnetic in 
principle. Then it is reasonable that the spin wave dispersion of $x = 0.12$ crystal 
is fairly isotropic with almost zero energy gap at the zone center \cite{r12}. The 
conductivity is apparently isotropic reflecting the isotropic ferromagnetic nature 
determined both bulk magnetic properties and spin dynamics, which is significant 
contrast to the character of $\rm LaMnO_3$ \cite{r13}.
\par
The most fascinating feature is that the resistivity jumps in enhanced manners under 
the applied field slightly above $T_L$ as shown in Fig. 3. It means that in this 
temperature range, $T_L<T<T_C$, there appears a striking phenomenon of the strong 
positive magneto-resistance effect in the CMR manganese oxides. Furthermore, the 
crystal undergoes the transition to a less distorted phase below $T_L$, accompanied 
with the metal-insulator transition behavior. It is highly possible, therefore, 
that the CO is established below $T_L$. Then we searched for the forbidden Bragg 
reflection of (2,0,1/2) below $T_L$ with a positive answer. Note that forbidden 
reflections of $(h,k,l+1/2)$ with integer $h,k,l$ are specified to the superlattice 
reflections due to the CO, which indicates that $\rm Mn^{4+}$ cations localize at the 
center surrounded 8 $\rm Mn^{3+}$ units in every other C layers [14]. The transition is 
definitely of the first order with distinct thermal hysteresis as well as the abrupt 
jump in intensity at $T_L$.  In order to confirm that this phase transition in the 
magnetic field is continued from the zero field, we performed the scans around 
both the fundamental Bragg reflection of (4,0,0) and the superlattice reflection 
of (2,0,1/2) changing the external magnetic field at several fixed temperatures 
between $T_L$ and $T_C$. Then we really observed that both structural transition and 
CO occur at the transition field assigned by the bulk measurements. Thus all the 
results concerning the phase transition are consistent with every respect. 
\section{Detection of orbital ordering by synchrotron x-ray scattering}
We now speculate that the drastic change in both magnetic properties and conductivity 
accompanied by the crystal distortion as described in the preceding section might 
be controlled by a "hidden" parameter of the "orbital" degree of freedom. The most 
decisive way to investigate the OO at this moment is applying the novel method of 
the synchrotron x-ray scattering technique established quite recently.
\par
By tuning the synchrotron radiation energies at the resonant transition energy 
between $1s$ core and $4p$ unoccupied electronic level corresponding to the energy 
of Mn K edge we could observe the orbital ordering structure, which was first 
demonstrated  for the CO and OO lro in the single layered  
$\rm La_{0.5}Sr_{1.5}MnO_4$ where 
$\rm Mn^{3+}$ and $\rm Mn^{4+}$ are mixed equally ($x=0.5$) 
by Murakami et al. \cite{r7}. Then the result 
of the OO driven by the {\it JT} interaction in 
$\rm LaMnO_3$ was also reported quite recently \cite{r15}. 
Here we briefly describe the principle of this method of the detection of 
the OO.
\par
The most important character of the synchrotron radiation source is not only the 
brightness of the coherent x-ray beam but the energy tunability by continuous 
scans of the monochromator. The polarization dependence of the anomalous scattering 
factor arises in the anisotropy of the charge, which may be determined either as 
the bond or orbital, which was first pointed out by V.E.Dmitrienko \cite{r16}. He defined 
the scattering of anomalous tensor of charge susceptibility (ATS scattering), since 
the anomalous scattering amplitudes of 
$\Delta f'(E)$ and $\Delta f'' (E)$ are written as the tensor, 
not the scaler for the normal charge scattering. Such a polarization dependent 
photo-absorption is readily known as the dichroism or the birefringence, in 
electromagnetic forces which rotates the polarization of the light. The essence 
of the polarization dependent phenomenon is the same as that of the anisotropy 
of the charge susceptibility caused through the resonant scattering, which is 
enhanced near the absorption edge. The ATS scattering enhanced at the energies 
near the K edge usually appears at the reflection forbidden by the atomic 
configuration of the crystal. Taking $\rm La_{0.5}Sr_{1.5}MnO_4$, for instance, in addition 
to a class of the forbidden $(h,k,0)$ reflections with $h$, $k$ of the half integer for 
the CO, the (3/4,3/4,0) is also the forbidden reflection which is only allowed by 
the OO of ($3z^2- r^2$) type in $e_g$ band of $\rm Mn^{3+}$. 
The ATS scattering due to the OO was 
clearly shown by the careful experiments; energy scans, polarization dependence 
($\sigma$ to $\pi$), ($\sigma$ to $\sigma$) or 
($\pi$ to $\pi$), the special scan with rotating the crystal around 
the scattering vector (azimuthal scan) and so forth. Polarization switching as well 
as the angle dependence in the azimuthal scan are only characterized for the ATS 
scattering. In other words, the amplitude of ATS scattering becomes a tensor rather 
than a scalar reflecting the anisotropy of the orbitals. 
\par
Then the  ATS scattering phenomenologically derived by Dmitrienko was clarified by 
taking the detailed mechanism of the polarization dependent synchrotron x-ray 
scattering into theoretical consideration \cite{r17}. The resonant elastic scattering near 
the K absorption edge is due to the elastic dipole transition. In other words, the 
synchrotron light excites electrons from $1s$ to $4p$ and then the resonant light is 
emitted resonating with the loss process of excited electrons. The formulation of 
the resonant x-ray scattering process [18] was extended to the present case of the 
anomalous process of the dipole transition. The anomalous scattering factor of 
$\Delta f'(E)$ and $\Delta f''(E)$ is written in the following formula, 
where the electronic system is excited 
from the initial state,  $|0 \rangle$ with energy $\varepsilon_0$ 
to the intermediate state, $|l \rangle$ with $\varepsilon_l$, 
and is finally relaxed to the final state, $| f \rangle$ with $\varepsilon_f$,      
\begin{equation}
\Delta f_{\alpha \beta} \propto { e^2 \over m c^2} 
\sum_l \Biggr \{ 
{   \langle f |j_{i \alpha} (-k') | l \rangle
    \langle l |j_{i \beta} ( k'') | 0 \rangle  \over 
    \varepsilon_0-\varepsilon_l-\omega_{k''} -i \delta }
 +
{   \langle f |j_{i \alpha} (k'') | l \rangle
    \langle l |j_{i \beta} (-k') | 0 \rangle  \over 
    \varepsilon_0-\varepsilon_l+\omega_{k''} -i \delta }    
 \Biggr \}
\ , \nonumber 
\label{eq:e1}
\end{equation}
\begin{equation}
j_{i \alpha} ={ e A_\alpha (\vec k) \over m} 
\sum_\sigma P_{i \alpha \sigma}^\dagger s_{i \sigma}
\ , \nonumber 
\label{eq:e2}
\end{equation}
\begin{eqnarray}
\alpha, \beta \ &:& \ {\rm polarization \ of \ photons} \\ \nonumber 
\omega_{k'(k'')} \ &:& \ {\rm incident\ (scattered)\  
photon \ energy \ with \ momentum} \ k'(k'') \\ \nonumber 
\delta \ &:& \ {\rm constant} \\ \nonumber 
A_\alpha(\vec k) \ &:& \ {\rm coupling \ constant} \ \ \ \nonumber 
(4p \ ({\rm operator} \ P^\dagger_{i \alpha \sigma} ) \rightarrow 1s (s_{i \sigma})) \ . 
\nonumber 
\end{eqnarray}
Here the real and imaginary parts of $\Delta f_{\alpha \beta}$ 
are respectively $\Delta f'(E)$ and $\Delta f''(E)$. 
Regarding to this formula, the dichroism or birefringence of the light 
occurs by the same type of  microscopic elastic dipole transition. Then 
a question arises why the ATS scattering occurs due to the $3d$ orbital order, 
or how the $3d$ orbital ordering reflects the anomalous scattering through $1s - 4p$ 
transition process. According to the detailed electronic structure of the $\rm MnO_6$ 
cluster, it was found that the $4p$ levels above the Fermi level lift the degeneracy 
mainly by the strong Coulomb interaction between $3d$ and $4p$ electrons in the same 
Mn cation, which should be defined the intra-atomic Coulomb interaction \cite{r17}. 
The $p_z$ orbital lifts from both $p_x$ and $p_y$, when $e_g$ orbitals are polarized in the 
basal plane. In fact, the calculation showed that the energy difference of the 
split  $4p$ levels due to the intra-atomic Coulomb interaction 
is $1.2 \pm 0.6 eV$, 
which is enough to induce the anomalous scattering in the resonant process. 
Then the scattering tensor can be calculated by assuming the $3d$ orbital ordering. 
The intensity of the forbidden interaction is also given in the following formula.
\begin{equation}
f= \pmatrix{
\Delta f_{xx} & 0 & 0 \cr 
0  & \Delta f_{yy} & 0 \cr 
0 & 0 & \Delta f_{zz} \cr  }
\end{equation}
\begin{equation}
I \propto | \Delta f_{xx} - \Delta f_{zz} | ^2
\end{equation}
The $\Delta f''_{xx(zz)}$ for either occupied  $d_{3z^2-r^2}$ 
or $d_{x^2-y^2}$ is the basis of the experimental 
observation of ATS scattering, which is the exactly same formulae as given 
phenomenologically. Hence, the anisotropy of the $d$ orbital reflects the tensor 
form in eq.(4) as well as the azimuthal angle dependence of the ATS scattering.
\par
So far, all the experimental results from the manganites revealed the appearance 
of the antiferro-type of the $d_{3z^2-r^2}$ OO where the z axis of polarization alternately 
directs in the $C$ plane as predicted. 
\par
Now, we understand completely the microscopic mechanism of the ATS scattering 
and recognize that this method gives a decisive experimental probe to investigate 
the OO, hopefully the orbital fluctuations by the inelastic scattering in near 
future. 
\par
We anticipated the OO by the insight from various experimental evidences 
described in the preceding section and  looked for the Bragg peak originated 
by the OO in the $\rm La_{0.88}Sr_{0.12}MnO_3$ single crystal. 
In fact, a distinct peak appears 
at (0,3,0) forbidden reflection below $T_L$ \cite{r9}. In addition of thermal scans, we 
have continued energy scans, azimuthal angle as well as polarization dependencies. 
All of the results show consistently that the 3D OO appears below $T_L$ as depicted 
in Fig.4. We emphasize here that the OO of the $\rm Mn^{3+}$ cation sites presented here 
occurs in the undistorted crystal of the pseudo cubic symmetry, in fact, the triclinic 
($P2$) symmetry according to the detailed powder x-ray diffraction measurement, 
whereas the previously observed or predicted OO arises in the {\it JT} distorted lattice. 
Judging from an important fact of the isotropic 3D ferromagnetic nature, the 
antiferro-type OO may not be the same pattern that of $\rm LaMnO_3$, but may be a unique 
orbital state of the hybridized orbitals of 
$d_{z^2-x^2(y^2-z^2)}$ and $d_{3x^2-r^2(3y^2-r^2)}$. 
predicted by the recent theories \cite{r19}.
\par
Now, we interpret this experimental fact in terms of the recent theoretical result 
based on the model Hamiltonian where both spin and orbital degrees of freedom are 
treated on an equal footing. The phase diagram is derived at $T=0$ with varying the 
exchange parameter of superexchange interaction of $t_{2g}$ band. This mean field calculation 
suggests that two kinds of ferromagnetic phase exist at the different carrier 
concentration, namely $x < 0.08$ and $x >0.42$ \cite{r19}. Then these two ferromagnetic phases 
are associated with two different orbital structures and there exists a  phase 
separated region between 2  phases in $0.08 < x <0.42$. This model calculation contains 
the common aspects with experimental evidence showing that the stabilization of the 
ferromagnetic insulator phase by both applying magnetic field and reducing 
temperature for the $x=0.12$ crystal : (1) both ferromagnetic ordering and antiferro-type 
orbital ordering are cooperatively stabilized in 3D manners, (2) the magnetic moment 
is apparently enlarged at the transition by changing dominant magnetic coupling from 
double exchange to superexchange interaction. Then more importantly,  this concept of 
the microscopic phase separation or the phase mixture  for the intermediate range, 
$T_L < T< T_C$, can be extended to the CMR effect occuring for $x > 0.25$ 
near below $T_C$, 
where the transition from the mixed phase with disordered orbitals which behaves 
like paramagnetic insulator  to the ferromagnetic metal. 
\section{Discussions and concluding remarks}
As described above, one of the essential results is the direct evidence of the 
OO in $\rm La_{0.88}Sr_{0.12}MnO_3$ single crystal with the same azimuthal angle dependence 
as that of the undoped $\rm LaMnO_3$. However it should be emphasized here the fact that 
the intensity appears only below $T_L$, where the notable {\it JT} distortion does not exist. 
Thus the antiferro-type OO is not the one associated with the {\it JT} lattice distortion. 
At the same time, the spin-wave dispersion relation as well as the electrical 
resistivity, magnetization in $\rm La_{0.88}Sr_{0.12}MnO_3$ 
are of all isotropic nature, which 
is fairly in contrast to that of undoped $ \rm LaMnO_3$ of 2D ferromagnetic character. 
Therefore, the orbital state in this crystal should be different from the ordering 
of $d_{3x^2-r^2}/d_{3y^2-r^2}$ in $\rm LaMnO_3$. 
The theoretically predicted OO is the hybridized orbitals 
as described in the preceding section. 
\par
Next, the intermediated state was found to be phase separated according to the 
precise structure refinement, though the phase separation does not start to coordinate 
with the phase transition at $T_H$ but grows around $305 K$. Since we could not observe 
both the OO and CO in the intermediate phase, though the scans were limited, we 
conjecture that the intermediate phase would not be uniform in many respects. 
We must consider more in the future experiments how the phase segregated structure 
in the intermediate temperatures can be relevant to the microscopically mixed phase 
as theory predicts. Since it is very essential to consider the inherent relation 
with the CMR effect, we leave this conclusion to the future investigations.
\par
As for the last point in the present paper, we propose a new concept of the anomalous 
phenomena appeared near the Mott transition, which should be common in the doped Mott 
insulator such as the high temperature superconducting copper oxides. It would contain 
a novel notion of the microscopic "phase separation" \cite{r20}. In the simple mean field 
view, the transition from insulator to metal, or Mott transition is considered to 
occur crossing the so called "critical" point, often called the "quantum critical"
 point for the doped cuprates. However, we don't think that the nature may be so 
simple. Furthermore the doped carriers cannot be uniformly distributed in the lattice 
space, in particular for the semi diluted concentration range of doped carriers. 
In this respect, the doped carriers might order at the special concentration such 
as 1/8 of the average uniform concentration of charges, which is very close to 
$ \rm La_{0.88}Sr_{0.12}MnO_3$. 
It also contains the new notion that the OO and CO is the result 
of the strong electron correlations, not driven by the {\it JT} distortion so far 
considered.  Though we need further experimental investigations, we observed for 
the first time that the metallic state in the intermediate phase shows the first 
order phase transition to the insulator by further cooling or applying the external 
magnetic field. Extending this scenario to the carrier rich concentration of 
$ \rm La_{1-x}Sr_xMnO_3$ where $x$ is around the CMR effect occurred, one might expect another 
phase transition from the mixed phase of the paramagnetic state showing the poor 
conductivity to the ferromagnetic phase with good conductive state by reducing the 
entropy either applying the magnetic field or cooling, which is defined the genuine 
CMR effect. Finally we would like to note that we do not have a concrete idea how 
the present model differs from the "polaron" model, but we believe that the role 
of the "orbital" is significant in the drastic phase transition occurred in 
$\rm La_{0.88}Sr_{0.12}MnO_3$. Nevertheless  the future experiments should clarify these two 
concepts.
\par
\medskip
\noindent
Acknowledgement
\par
Authors acknowledge dedicating assistance of A.Nishizawa and M.Onodera in the  
single crystal preparation of manganese oxides. They also thank Y.Tokura, 
G.Shirane, N.Nagaosa, D.E.Cox, M.Blume for their stimulating discussions 
throughout the present work. The work has partly been supported by a Grant 
in Aid for Scientific Research from the Ministry of Education, Science, 
Sports and Culture of Japan in addition to the Core Research for Evolutional 
Science and Technology (CREST) by the Japan Science Technology Corporation.

\vfill
\eject
\noindent
Figure captions
\par
\medskip
\noindent
Fig.1: Magnetic and structural phase diagram for $\rm La_{1-x}Sr_xMnO_3$ ($0 < x < 0.3$ ) 
determined by neutron diffraction data depicted at the left hand side. 
PM, CAF and FM are, respectively, the symbols of paramagneic, canted
antiferromagnetic and ferromagnetic phases. $O^\ast$ and $O'$ are the phases with
two different orthorhombic structure. O* was defined to be pseudo cubic in
the original litarature \cite{r8}.
\par
\noindent
Fig.2: Sequential phase transitions observed by neutron diffraction for
$\rm La_{0.88}Sr_{0.12}MnO_3$ single crystal. Upper panel shows lattice vector of
fundamental reflections and the middle shows peak intensities. The bottom
is peak intensities of the ferromagnetic Bragg reflection which starts to
grow belo $T_C = 170 K$. The phase transition at both $T_L = 145 K$ and 
$T_H = 291$
K is of the first order \cite{r9}. 
\par
\noindent
Fig.3: Upper panel shows magnetization curves measured in pulsed magnetic
fields for $\rm La_{0.88}Sr_{0.12}MnO_3$ single crystal.  The vertical origin is shifted for
the convenience by $-0.4mB$ each per $5 K$ shift. Lower panel is 
magneto-resistance at various temperatures around our interests \cite{r10}.
\par
\noindent
Fig.4: ATS scattering from $\rm La_{0.88}Sr_{0.12}MnO_3$ single crystal. Upper panel shows the
data of energy scan. Distinct peak at (0 3 0) appears at 6.552 keV, where
absorption reaches maximum.  The middle shows the azimuthal angle scan
and the bottom shows thermal vstistion of the peak intensities \cite{r9}.	
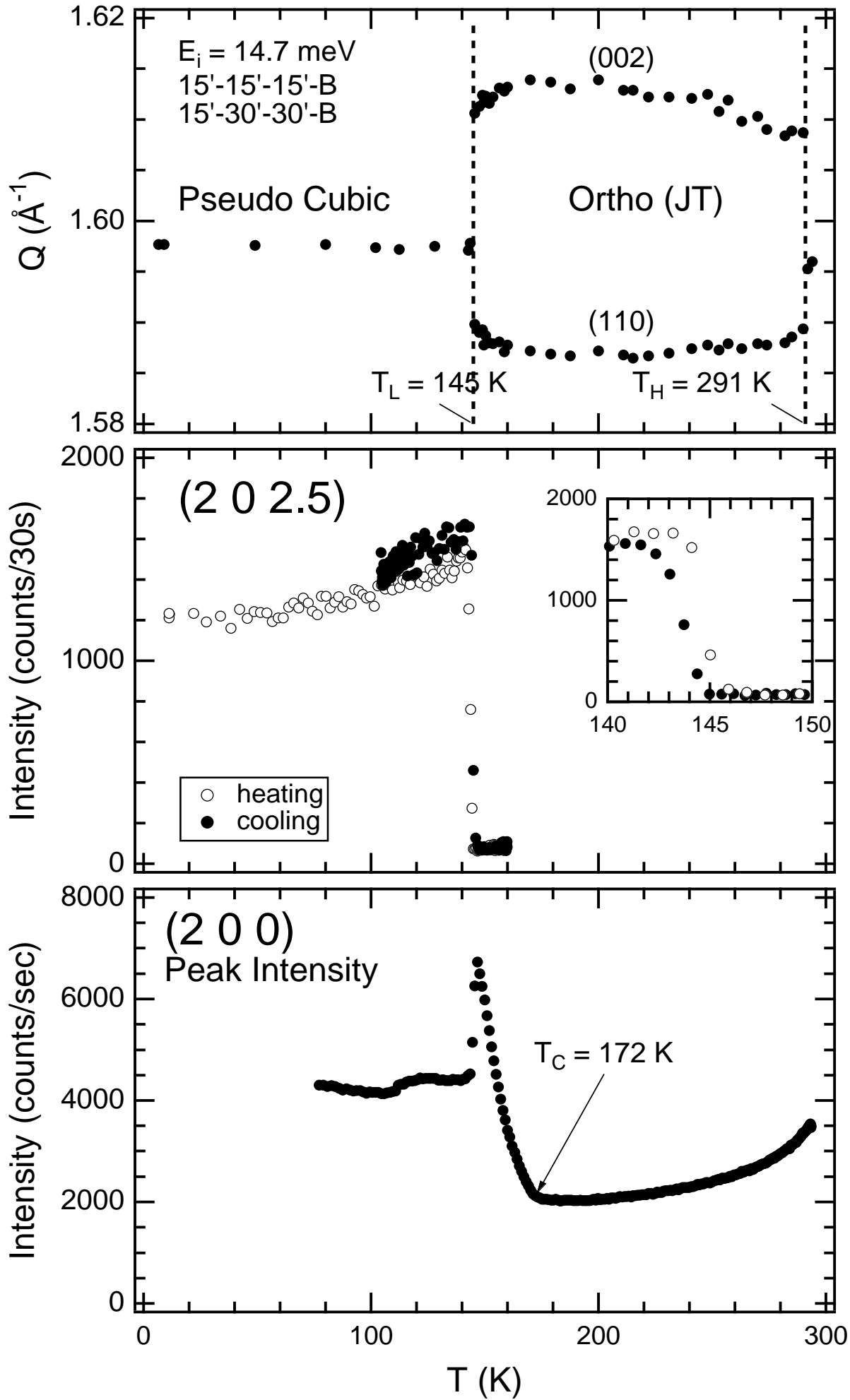
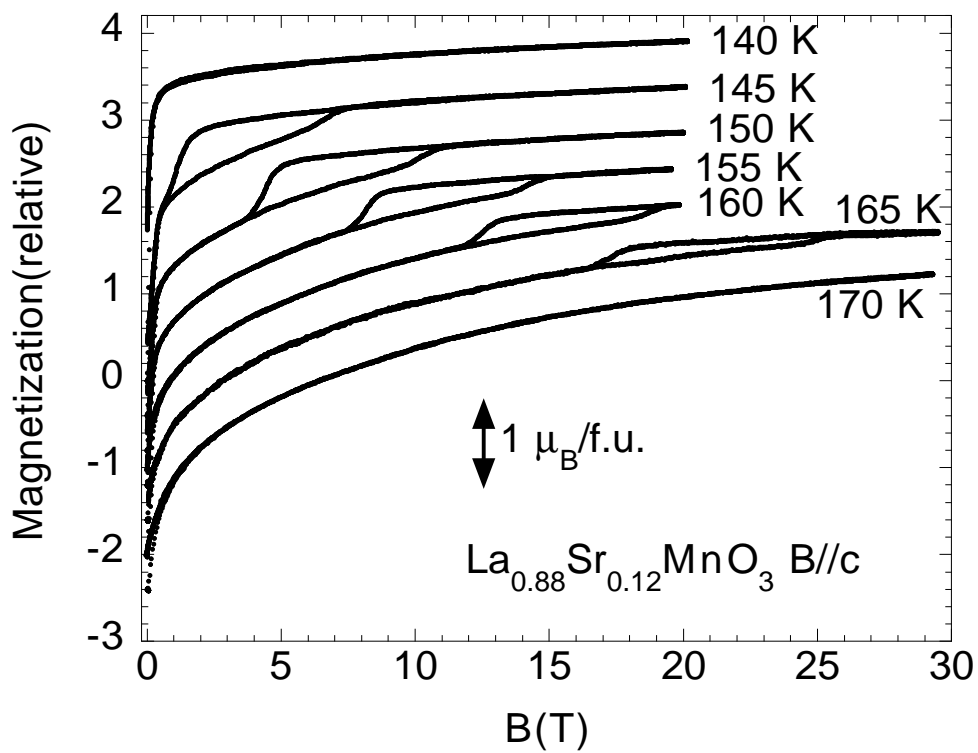
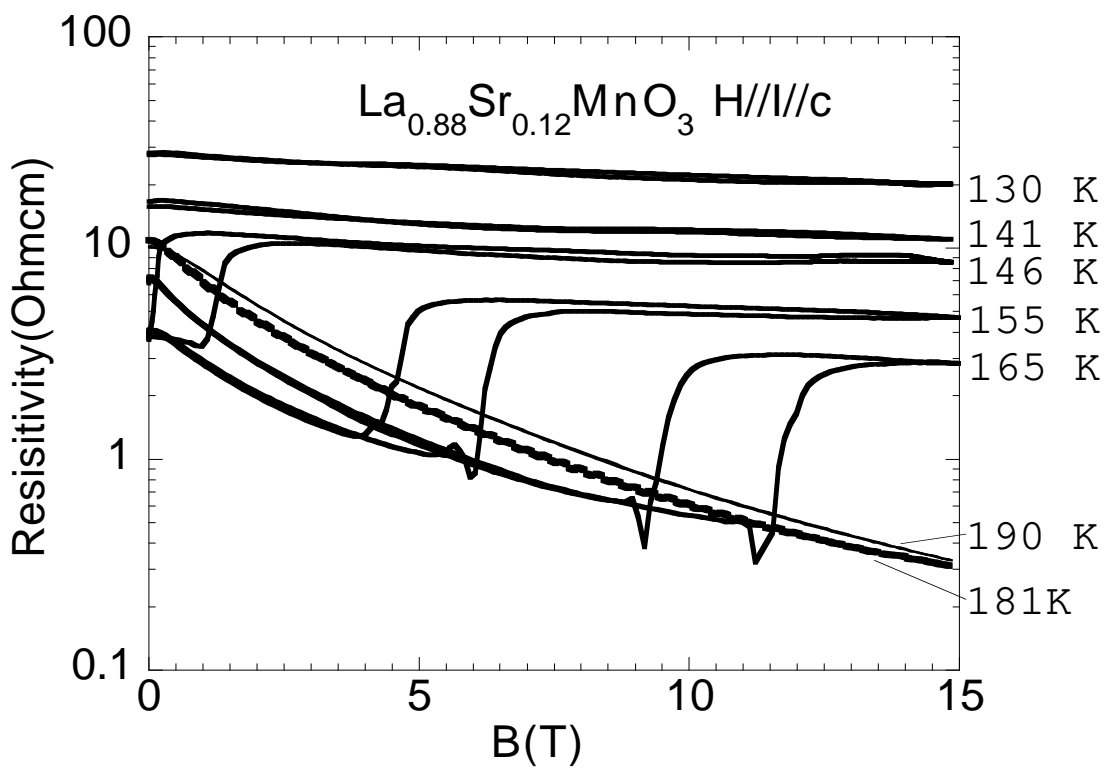
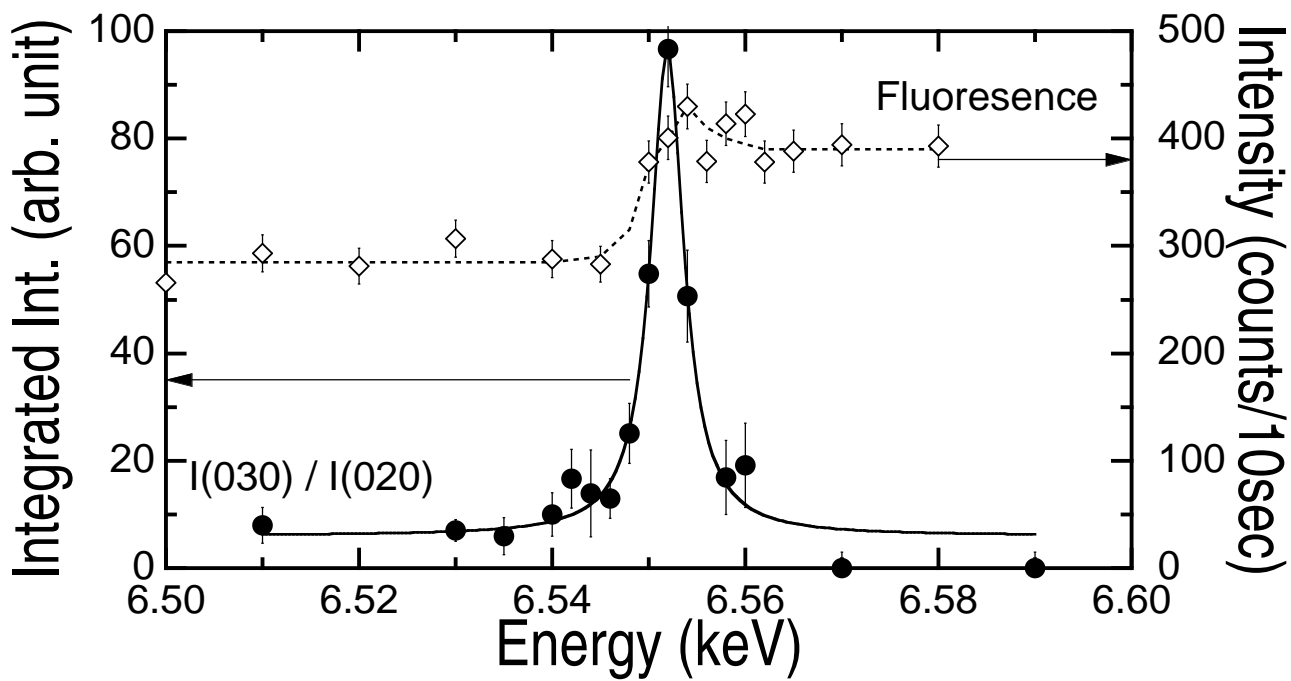
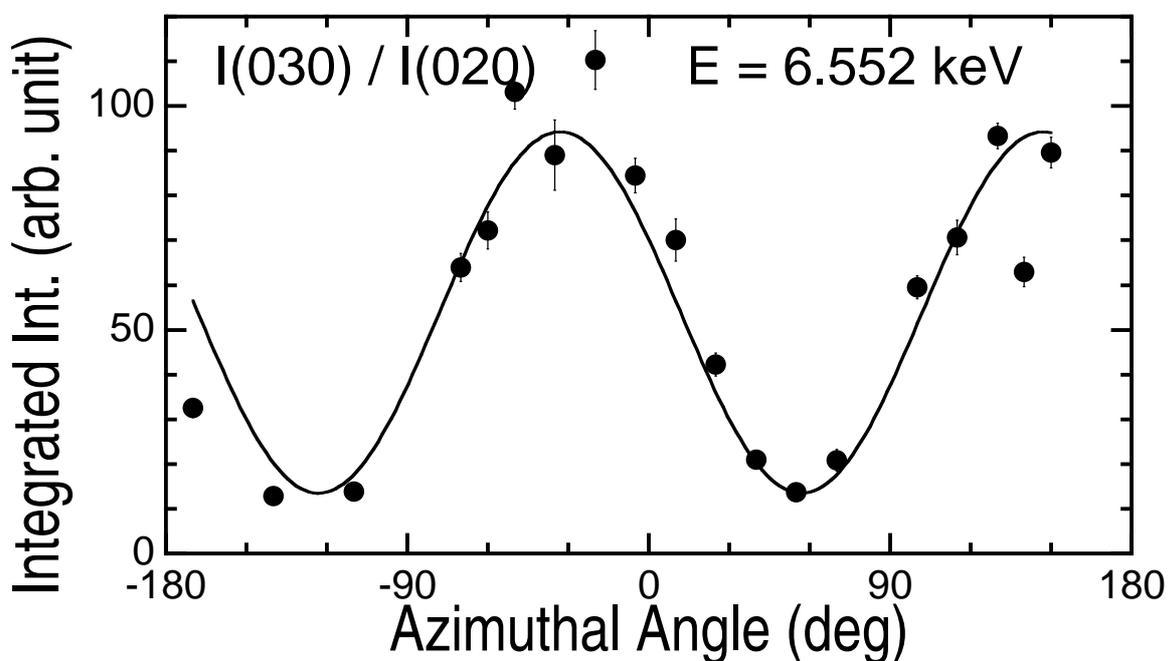
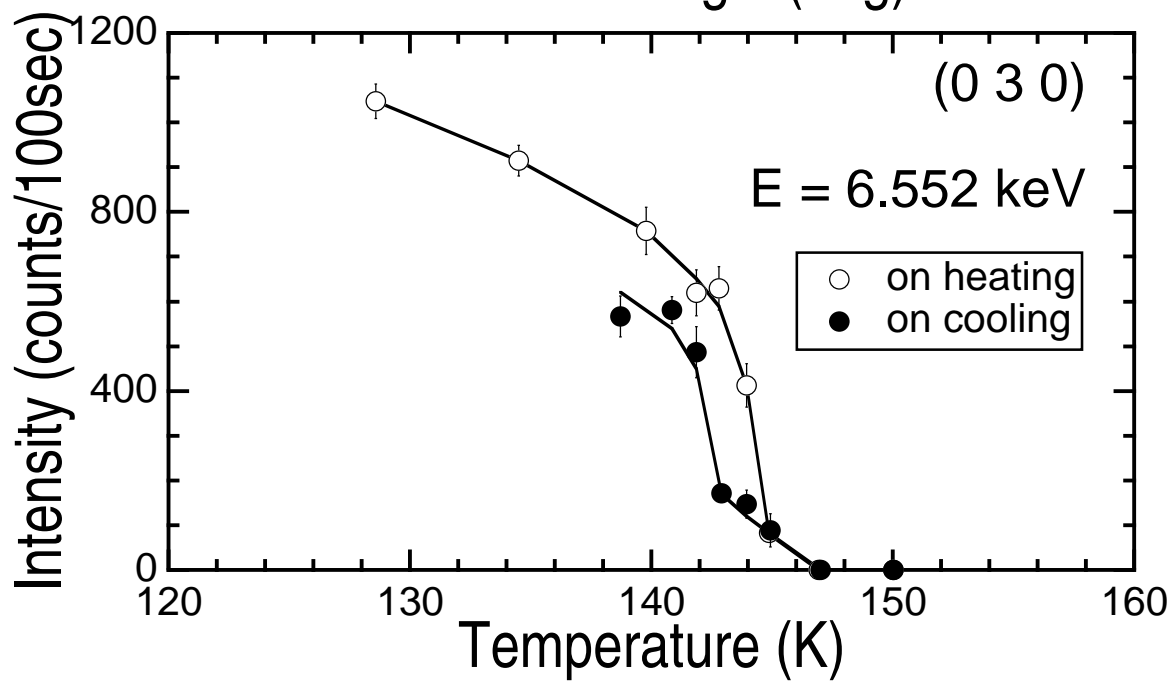
\end{document}